\documentclass[USenglish]{article}	

\usepackage[big,online]{dgruyter}	
\usepackage{lmodern} 
\usepackage{microtype}
\usepackage[numbers,square,sort&compress]{natbib}

\usepackage{pstricks}
\usepackage{arydshln}
\usepackage{graphicx}

\begin{document}

\title{Personnel-adjustment for home run park effects in Major
League Baseball}
\runningtitle{Personnel-adjusted park effects}

\author*[1]{Jason A. Osborne}
\author[2]{Richard A. Levine}
\runningauthor{Osborne and Levine}
\affil[1]{\protect\raggedright
North Carolina State University, Department of Statistics, Raleigh, NC, USA, email: jaosborn@ncsu.edu}
\affil[2]{\protect\raggedright
San Diego State University, Department of Math and Statistics, San Diego, CA,
USA, e-mail: rlevine@sdsu.ed}

\abstract{
In Major League Baseball, every ballpark is different, with
different dimensions and climates. 
These differences make some ballparks more conducive to hitting
home runs than others.
Several factors conspire to make estimation of
these differences challenging.  Home runs are relatively rare, occurring in
roughly 3\% of plate appearances.  
The quality of personnel and the frequency of batter-pitcher handedness
combinations that appear in the thirty ballparks vary
considerably. 
Because of asymmetries, effects due to ballpark can depend strongly on
hitter handedness.
We consider generalized linear mixed effects models 
based on the Poisson distribution for home runs. 
We use as our observational unit the 
combination of game and handedness-matchup.  
Our model allows for four 
theoretical mean home run 
frequency functions for each ballpark.
We control for variation in personnel across games by constructing
``elsewhere'' measures of batter ability to hit home runs and pitcher tendency
to give them up, 
using data from parks other than the one in which the response is observed. 
We analyze 13 seasons of data and find
that the estimated home run frequencies
adjusted to average personnel are substantially different from observed home
run frequencies, leading to considerably
different ballpark rankings than
often appear in the media.
}


\maketitle

\section{Introduction} 
\subsection{Motivation}

One of the enjoyable aspects of Major League Baseball (MLB) is that
every ballpark is different, each with its own geometry.
It is easier for left-handed batters (LHB) to hit home runs 
in some ballparks and easier for right-handed batters (RHB) in others.  
The chance that a plate appearance (PA)
results in a home run depends on many factors whose effects
can be modeled with the wealth of data available to the modern
data analyst.  

Much work has been done to quantify differences among ballparks,
much of it in popular media, and some in scholarly literature,
but little of this work involves directly modeling effects of
individual batters and pitchers.  
Our approach involves construction of measures which quantify individual 
batter abilities to hit and pitcher tendencies to allow home runs.  
These explanatory variables
are based on plate appearances observed at other parks, avoiding
the problem of directly using the same data to estimate park effects 
and player abilities simultaneously.  Further, while recent progress
has been made to separately quantify park effects for LHB and RHB,
most approaches typically average over pitcher-handedness, despite the 
common knowledge that RHBs tend to perform better against left-handed
pitchers (LHP) than against right-handed pitchers (RHP), and conversely
for LHBs.  The frequencies of pitcher handedness in ballparks can vary
dramatically in the same way that batter handedness does, so it makes
sense to estimate home run frequency for fixed pitcher-handedness 
and also after sensibly averaging over it.

In a chapter devoted to estimation of park factors, \cite{abdwr2} observed
a limitation of many approaches to the problem. (The parenthetical
remarks in the following quoted sentence are the authors' not 
ours): ``Finally, the proposed park factors  (as well as most published 
versions of park factors) essentially ignore the players involved in 
each event (in this case the batter and the pitcher).''  This 
observation emphasizes the need for models to include effects for 
individual players and motivates our approach.  Furthermore, existing methods tend to stop at some numerical summary of the degree to which a park favors offense, stopping short of uncertainty quantification, an important issue when estimates are often based on 1-3 years worth of data. Our estimates are simple functions of model parameters so their standard errors are readily available.

After a literature review, Section 2 outlines the methods, including 
an articulation of the statistical model and a description of how
covariates in the model are constructed.  The data analyses and
statistical inferences are in Section 3, with estimates of both 
matchup-specific estimates
of home run frequencies and sensibly averaged marginal ballpark
means.  Other analyses include quantification of the degree
to which residual variance is reduced when going from subset
models to a full model and an investigation of the degree to 
which MLB division explains variability in HRs.   Section 4 provides
a summary and suggestions for further research.

\subsection{Literature Review}
A measure of home run friendliness popularized by the ESPN network, known as 
the home run park 
factor (HRPF),
can be computed for each ballpark by comparing home runs 
hit and allowed at home by 
the team that plays their home games at the ballpark with 
home runs hit and allowed by that same team at other parks:
$$ HRPF = \frac{(HR_{hit,Home}+ HR_{all,Home})/G_{Home}}{(HR_{hit,Rd}+ HR_{all,Rd})/G_{Rd}}.$$
Positive values suggest home run friendliness of the ballpark.  The idea behind
the metric is to control for variability due to team.

\cite{acharya} show that when this measure is used to assess runs 
scored across ballparks, it is biased due to scheduling imbalance.  
Separate models are developed for the American and National Leagues
that include factorial effects for park.  The model also includes
factorial effects for offensive and defensive team, though it does
not model contributions of individual players 
nor their handedness.
Many managers
make use of the ``platoon'' to take advantage of matchups, so
that lineups vary substantially from one game to the next.
In a detailed investigation of handedness effects on home runs 
and many other offensive measures, \cite{Chance} analyze a 
century's worth of data and break matchups down further by considering batter 
throwing hand to obtain eight matchup combinations.
Neither linear models with factorial effects for teams and
parks, nor the HRPF take variable lineups 
or handedness combination into account.

When modeling the chance of a base hit in a PA, \cite{alceo}
compare 48 models, including generalized linear models, random
forest and neural networks.  
The HRPF 
is 
selected as an explanatory variable in many of these.
In a study of air density effects on distance travelled by batted
balls \citep{bahill}, 
the following factors are identified, in decreasing order of variance 
explained: altitude, temperature,
barometric pressure and humidity.
A comparison of home ballpark advantage across the 2019
and 2020 (Covid) seasons, to investigate the possible effect of fan absence
\citep{losak}, found no difference.

\cite{abdwr2} develop one approach to estimate park effects
using games as the observational unit and another based on
batted balls.
A model focussing
on home runs 
that takes batted balls as the observational unit is also developed.
The authors demonstrate how these park factors can be used to adjust
estimates of 
home run
frequencies for individual players to what they would
hit in a neutral ballpark.  
Many articles addressing park effects appear in the Sabermetrics 
literature.  
\cite{fox} studies atmospherics, with several important findings.
Pitches decelerate more in evening games, with observed correlation
between temperature and deceleration of $r=0.61$.  
Average relative decreases in pitch velocity ranged from 5\% (Comerica
Park in Detroit) to 11\% (T-Mobile Park in Seattle.)

\cite*{click} studies variation in park effect estimators over time.
\cite{nesbit} study the roll of park factors on attendance in baseball,
and do not find an effect.
\cite{perry1} estimates park effects separately for
LHB and RHB and discusses lineup variation and pitcher
handedness.
\cite{rybarczyk} simulates batted ball data and uses trajectories to 
 estimate park factors specific to five outfield sections: 
left, left-center, center, right-center and right field.
The following factors are identified as likely to have an effect 
on home run 
frequency:
 atmospherics, roster, roster changes, ball 
characteristics (dry $>$ humid), unbalanced schedule, interleague play, 
and weather.  Highlighting the considerable variability from one year to
the next, \cite{rybarczyk} notes that the HRPF 
for Chase Field in Arizona rose from 48 in 
2002 to 116 in 2003.

Justifiably, there has been considerable research into atmospheric
effects on batted balls and how the weather can lead to 
differences among ballparks.  There has been less work done to 
account for differences in personnel, and even less in articulating how
handedness affects home runs 
differently in different ballparks.
Part of the difficulty in doing so is that assessment of player 
abilities is based on the same data used to differentiate 
ballparks.  We develop generalized linear models based on the
Poisson distribution that allow for batter and pitcher effects that 
vary across the four combinations of batter and pitcher handedness.

\section{Methods}
\subsection{Data}
We analyze data from MLB seasons 2010-2023, excluding the partial
2020 season.  Play-by-play accounts of games from these seasons 
were collected using Retrosheet, a non-profit organization with a wealth of
information about baseball from the last century \citep{Retrosheet}.
Home runs are aggregated over each game, separately
for each of the four combinations of batter and pitcher handedness:
with left-handed batter vs left-handed pitcher denoted LL,
left-handed batter vs right-handed pitcher denoted LR,
right-handed batter vs left-handed pitcher denoted RL, and
right-handed batter vs right-handed pitcher denoted RR.

The unit of observation
is the combination of game and matchup.  
The total number of observations analyzed here 
is $N=$123,886, involving 2,451,105 PAs.
If there were four matchups
observed for each of 81 games in each of 30 parks over these 13 seasons,
there would be $81 \times 30 \times 13 \times 4 = 126,360$ observations.  
However, games played
by the then Florida Marlins, in Joe Robbie Stadium in 2010 and 2011
were excluded from consideration, as were any other games played outside
of the current 30 ballparks.  Additionally, only $87\%$ of all games 
included plate appearances from the full complement of all four 
matchups.   (See Table \ref{matchupfrqs} for a breakdown of all matchup frequencies.)

\begin{table} [!ht]
\centering
\caption{Matchup frequencies of batter and pitcher handedness.}
\begin{tabular}{c|c|c|c} 
Batter & \multicolumn{2}{c|}{Pitcher Hand} & \\ \cline{2-3} 
Hand   & LHP     & RHP     & Total \\ \hline
LHB    & 193626  & 834352  & 1027978 \\ 
       & (0.079) & (0.340) & (0.419) \\  \hline
RHB    & 483065  & 940062  & 1423127\\ 
       & (0.197) & (0.384) & (0.581) \\ \hline
Total  & 676691  & 1774414 & 2451105 \\ 
       & (0.276) & (0.724) &  \\ \hline
\end{tabular}
\label{matchupfrqs} 
\end{table}

\subsection{Model}
We adopt a generalized linear mixed effects model 
with a canonical log link function.  The factors included
in the model are listed below and are followed by description and
justification:

\begin{itemize}
\item matchup: combination of batter and pitcher handedness,
\item ballpark-by-matchup interaction,
\item aggregated measures of batter and pitcher HR 
propensity  (two ``personnel covariates''),
\item season (using random effects).
\end{itemize}
It is well-known in baseball that the 
batter-pitcher handedness combinations that are generally most conducive to offense
are those where the handedness of the batter and pitcher
do not agree.  Table~\ref{hrfmatchup} shows the relative
frequencies of home runs per PA for the four matchups.  In light
of this variability, the model includes matchup effects.

\begin{table} [!ht]
\centering
\caption{Home run frequencies by matchup.}
\begin{tabular}{c|c|c|c} 
Batter & \multicolumn{2}{c|}{Pitcher Hand} & \\ \cline{2-3}
Hand   & LHP   & RHP & Total \\ \hline
LHB    & $4267/193626 = 0.022$  & $24418/834352 = 0.029$  & $28685/1027978 = 0.028$  \\ \hline
RHB    & $14837/483065 = 0.031$  & $27472/940062 = 0.029$ & $42309/1423127 = 0.030$ \\ \hline
Total  & $19104/676691  = 0.028$ & $51890/1774414 = 0.029$ & $70994/2451105 = 0.029$ \\ \hline
\end{tabular}
\label{hrfmatchup}
\end{table}

Given this evidence that home run intensity varies across 
matchups, it is important to note that the distribution
of PAs across matchups itself varies considerably across
ballparks.  Consider Progressive Field, home to the
Cleveland Guardians.
While league-wide relative frequency
of PAs by left-handed batters is $0.419$,
the figure in Cleveland is quite different at $0.514$.
Since more than half of all PAs in Cleveland
are by LHB (see Table \ref{progfield}), it is unsurprising 
that more HRs per game by LHB against RHB have been observed 
there than at any other ballpark, even Yankee
Stadium, famously friendly to left-handed batters.   
To account for this variation
in matchup frequency across parks, we allow four parameters,
one for each matchup, for each ballpark, for a total of
120 parameters of primary interest in the model.  
This leads to separate and very different estimated rankings of 
ballparks for each of the four matchups.

\begin{table} [!ht]
\centering
\caption{Matchup frequencies observed at 
Cleveland's Progressive Field ($\sim 80000$ PAs).  LHP, RHP
abbreviate left- and right-handed pitcher, respectively,
and LHB and RHB abbreviate batter handedness.}
\begin{tabular}{c|c|c|c} 
& LHP & RHP & Total \\ \hline
LHB & 0.073 & 0.440 & 0.514 \\
RHB & 0.145 & 0.341 & 0.486 \\ \hline
Total & 0.218 & 0.782 &  \\ \hline
\end{tabular} \\ 
\label{progfield}
\end{table}

Beyond matchup frequency, we also seek to control for variability 
due to quality of personnel having plate appearances at the
various ballparks. Certain high-revenue teams, like the
Los Angeles Dodgers and New York Yankees, dedicate 
more resources to signing free agent batters than 
others. 
In 2023, the team hitting the most HRs hit more than twice as many as the
team hitting the fewest.  Further, there can be considerable game-to-game
variation in lineups for a given team.  With respect to allowing home runs,
pitchers appearing in ballparks can also vary from team to team,
game to game
and within a game.
Controlling 
for these personnel effects offers the potential to explain
variability in home run 
hitting.  For each observational unit (combination
of game and matchup), we construct ``personnel covariates'' that quantify the home run
hitting ability of batters and home run allowing tendency of 
pitchers for all 
PAs for a particular matchup in a game.  We use the term covariate since
this is not under any kind of control, and we will construct adjusted ballpark
factors by estimating the mean number of home runs in a game played there at 
average values of this covariate.  There are two such covariates for each
observation, one for batters and one for pitchers.

In an attempt to normalize home-run hitting across ballparks, MLB
has been experimenting with several approaches, not limited
to keeping baseballs in humidors prior to games.  For whatever
reason, home run intensities vary over time.  
Over the 13 non-Covid year period of observation considered here,
the lowest HR intensity occurred in 2014 at $1.72 \ HR/g$ 
but five years later that figure rose to $2.78 \ HR/g$, an 
increase of $62\%$. 
Rather than attempting to model non-monotonic 
local trends over time, possibly varying across ballpark, we include 
an additive random effect for season.

\subsection{Elsewhere measures of Batter and Pitcher quality} 

As observed by 
\cite{acharya}, the problem of identifying who 
the productive batters are is itself made difficult by
ballpark biases. 
Furthermore,
the
data used to assess 
player 
quality is the same data we seek to
use to assess the home run friendliness of the ballparks.
To get around this, we propose ``elsewhere'' measures of hitter
quality.  For a particular PA, we construct a measure
of quality for the batter, $q^B$, using the relative frequency with
which he has hit a HR in {\em all other ballparks} excluding
the one in which the PA is taking place.  We construct
a measure for the pitcher, $q^P$ in a similar fashion.
Both measures are specific to matchup.

As an illustration, consider an example from the 2021 home opener for the
Colorado Rockies at Coors Field, facing the Los Angeles Dodgers
who had their ace LHP Clayton Kershaw starting the game.
For the first plate appearance in that game by the Rockies shortstop
Trevor Story, a RHB, we observe that in the 562 PAs by Story
against LHPs away from Coors Field he has hit 25 HRs.  Our
``elsewhere'' measure for that PA is then $q^B=\frac{25}{562}=0.044$.
Similarly for the pitcher Kershaw, who has pitched in 7105 PAs 
against RHBs away from Coors Field and given up 150 HRs, we construct 
$q^P=\frac{150}{7105}=0.021$, an especially low rate. 
This is not a perfect measure as it will be heavily influenced by 
Kershaw's home
ballpark of Dodger's stadium, but it does get around
the problem of using the same data to assess player
quality as are used to assess ballpark HR friendliness.
After computing these batter and pitcher quality measures, we aggregate
them by matchup for all PAs in a given game to obtain ``z-variables''
to be used in the right-hand side of our generalized linear model:
$Z^B_i = \sum_{PA \in i} q^B$ and $Z^P_i= \sum_{PA \in i} q^P$
where $i$ is an index for all game-matchup combinations in the data set.
For a game-matchup observation $i$ where there were many good batters 
and many bad pitchers, we would see large values of $Z^B_i$ and $Z^P_i$.
Lastly, we include variables for matchup-specific ballpark effects (four
per ballpark) using 30 indicator variables.

The total number of observed HR counts for game-matchups was 
$N=$123,886.  If the HR totals for these are denoted $Y_1,\ldots,Y_{123886}$
our model assumes these to be Poisson random variables
with log-means given by the following regression equation:

\begin{eqnarray*}
\lefteqn{\log \lambda_i(X^S_i,Z^B_i,Z^P_i,X^{LHB}_i,X^{RHB}_i,X^{LHP}_i,X^{RHP}_i,X^P_i) = } \\ 
&   & \beta_0 +  \beta^B Z^B_i + \beta^P Z^P_i + \sum_{s=1}^{13} S_s X^{S}_{i,s}  +\\ 
&   & \sum_{p=1}^{30} \left(\beta^{LL}_p X^{LHB}_i X^{LHP}_i + \beta^{LR}_p X^{LHB}_i X^{RHP}_i + \beta^{RL}_p X^{RHB}_i X^{LHP}_i + \beta^{RR}_p X^{RHB}_i X^{RHP}_i\right) X^p_i. 
\end{eqnarray*}
Here, $X^S_{i,s}$ are indicator variables for seasons, $s=1,\ldots,13$.
$p$ is an index for ballparks.  
$Z^B_i$ and $Z^P_i$ are the elsewhere measures of batter and pitcher tendency 
with regard to hitting and allowing HRs, respectively.  The last summation
in the model is over 30 ballparks, with ballpark indicator variables $X_i^1,X_i^2,\ldots,X_i^{30}$.  
The left-handedness and right-handedness indicators for 
batters are denoted $X^{LHB}_i$ and $X^{RHB}_i$ and similarly for the indicators for pitchers, $X^{LHP}_i$ and $X^{RHP}_i$. 
Having defined these explanatory variables, we define
$S_s$ for $s=1,\ldots,13$ as the 
season effects, assumed to be
a random sample from a mean zero normal population with variance component $\sigma_S^2$.  
$\beta^B$ and $\beta^P$ are the personnel effects quantifying the
degree to which batters and pitchers, respectively, contribute to home run intensity.
Lastly the parameters of primary interest are the 120 ballpark parameters
for the four matchups, $\beta_p^{LL},\beta_p^{LR},\beta_p^{RL},\beta_p^{RR}$
for ballparks $p=1,\ldots,30$.

When a data frame is suitably formatted to fit the model with statistical
software it might have one row for each game-matchup, 
a column with the response, the number of home runs hit for that game-matchup,
and columns for season, personnel variables, ballpark, and the indicators for
batter- and pitcher-handedness. 
The annotated 
data in Table \ref{datatable}
gives an illustrated example
of the eight observations from the first two games played in 
Coors Field in 2021.

\begin{table}
\centering
\caption{Data frame for eight observations from the first two games at Coors Field in 2021.  {\tt zB} 
is the matchup-specific aggregate of the rates at which home runs were hit by 
all batters with plate appearances in the game.  {\tt zP} is a similar
measure for pitchers.  {\tt pa} is the matchup-specific frequency of plate 
appearances.  The value with asterisk includes contributions to the
personnel measure from the batter Trevor Story when facing left-handed pitchers and
the pitcher Clayton Kershaw when facing right-handed batters, with these
figures computed from data away from Coors Field.  The small value of 
{\tt zP} despite the large number of plate appearances is due to the
low rate at which Kershaw gives up home runs when facing right-handed batters.
The observed number of home runs is in the {\tt hrsum} column. }
\begin{verbatim}
  GAME_ID       hrsum       zB       zP    park    bh    ph    pa   season

COL202104010      0      0.138    0.187    COL     L     L      9     2021
COL202104010      0      0.994    0.725    COL     L     R     26     2021
COL202104010      0     *0.537*  *0.489*   COL     R     L     24     2021
COL202104010      0      0.808    0.832    COL     R     R     26     2021
COL202104020      2      0.445    0.448    COL     L     L     12     2021
COL202104020      2      1.066    0.970    COL     L     R     30     2021
COL202104020      0      0.245    0.107    COL     R     L      8     2021
COL202104020      0      1.023    0.888    COL     R     R     34     2021
\end{verbatim}    
\label{datatable}
\end{table}

Recall that for each matchup, these observations are aggregates over 
all PAs in the game.  The asterisks in 
Table \ref{datatable}
emphasize that these $z$-variables
include the example elsewhere measures we constructed for Trevor
Story and Clayton Kershaw, along with all other RL matchups in
the game.

The fitted values from the model are estimated home run expectancies
for the game-matchup combinations, 
$\widehat\lambda_1,\ldots,\widehat\lambda_{123886}$.  
To assess the fit of the model, we round these expectancies to the nearest
0.2 and compare the observed distribution of HR counts for the corresponding
game-matchups with a Poisson distribution.  For example, consider the subsample
of $N=3132$ such observations for which $\widehat\lambda_i$ rounds to 1.4 expected
HRs.  Table \ref{HoaglinTukeyTable}
gives the observed counts $n_k$ of
game-matchups in which $k=0,1,\ldots,7$ HRs were hit.

\begin{table} [!ht]
\centering
\renewcommand{\arraystretch}{1.1}
\caption{Frequencies of home run counts in the $3132$ 
game-matchups where the estimated mean is $\widehat\lambda=1.4$.  
Columns are, respectively, observed frequencies, $n_k$,  observed relative
frequencies, $n_k/3132$, fitted frequencies, Hoaglin-Tukey transformed
relative frequencies, $\phi(k)$.}
\begin{tabular}{c|c|c|c|c} 
Home & Observed & Relative  & Fitted     & Hoaglin-Tukey \\
Runs & Frequency& Frequency & Frequency  & $\phi(k)=$\\ 
$k$ & $n_k$ & $n_k/3132$ & $e^{-\widehat\lambda}\widehat\lambda^k/k!$ & $\log(k!n_k/3132)$ \\ \hline

 0      &           816       &        0.26       &       0.25      &      -1.35 \\ 
 1      &          1115       &        0.36       &       0.35      &      -1.03 \\ 
 2      &           704       &        0.22       &       0.24      &      -0.80 \\
 3      &           331       &        0.11       &       0.11      &      -0.46 \\
 4      &           131       &        0.04       &       0.04      &       0.00 \\
 5      &            24       &        0.01       &       0.01      &      -0.08 \\
 6      &            10       &        0.00       &       0.00      &       0.83 \\
 7      &             1       &        0.00       &       0.00      &       0.48 \\ \hline
\end{tabular}
\label{HoaglinTukeyTable}
\end{table}

There were $n_1=816$
game-matchups where no HRs were hit, constituting $26\%$ of the
subsample.  The Poisson model yields an expected, fitted fraction of 
$\widehat{P}(N=0;\lambda=1.4) = e^{-1.4}(1.4)^{0}/0!=0.25$.  This
fitted relative frequency appears along with those for $k=1,2,\ldots,7$
in the fourth column of Table \ref{HoaglinTukeyTable}.  Agreement of the empirical
and fitted frequencies in the third and fourth columns of the table
is strong.  
Figure 1 plots these two columns; the estimated Poisson 
probability mass function together with the empirical relative 
frequencies of $k$ HRs with $k=0,\ldots,7$. For the game-matchups
with $\widehat\lambda=1.4$ considered here, there does not appear
to be much evidence of lack of fit of the Poisson model.

\begin{center}
Figure 1 goes about here.
\end{center}

We extend the examination to game-matchups with other estimated
home run expectancies using Poissonness plots \citep{hoaglin}.
If the probability mass function is multiplied by $k!$ and
log-transformed, then the resulting 
expression, $-\lambda + k\log(\lambda)$, is
linear in $k$.  If the same transformation is applied to the
empirical relative frequencies, $\phi(k)=\log(k!n_k/N)$, it should be
approximately linear with intercept $-\lambda$ and 
slope $\log(\lambda)$, under the Poisson model.
Inspection of the last column of Table \ref{HoaglinTukeyTable} shows a 
roughly linear increase with $k$ up until $k \geq 5$.  
For HR totals where observed counts are smaller, the 
relative variation of $\phi(k)=\log(k!n_k/N)$ function is greater.
The plot of $\phi(k)$ against $k$ for these 3132 game-matchups
can be found in Figure 2, along with similar plots
corresponding to other values of $\widehat\lambda$.  As these plots
are inspected left-to-right, top-to-bottom, the increase in slopes
is roughly linear, going from negative to positive with $\log(\lambda)$.
None
provide much evidence of lack-of-fit of the Poisson model.

\begin{center}
Figure 2 goes about here.
\end{center}

\section{Data Analysis}

Upon substitution of the location parameter estimate, the following 
equation for the log of the mean home runs hit in game-matchup combination $i$
is obtained:
\begin{eqnarray*}
\lefteqn{\log(\widehat\lambda_i) = } \\
& & -1.3671 + 1.0127 \ Z^B_i + 0.3841 \ Z^P_i +  \\
& & \sum_{p=1}^{30} \left(\widehat\beta^{LL}_p X^{LHB}_i X^{LHP}_i + \widehat\beta^{LR}_p X^{LHB}_i X^{RHP}_i + \widehat\beta^{RL}_p X^{RHB}_i X^{LHP}_i + \widehat\beta^{RR}_p X^{RHB}_i X^{RHP}_i\right) X^p_i. 
\end{eqnarray*}
The estimates for the 120 ballpark effects parameters, $\widehat\beta^{LL}_p,\widehat\beta^{LR}_p,\widehat\beta^{RL}_p,\widehat\beta^{RR}_p$
for $p=1,\ldots,30,$ are given in Table \ref{parms}.
The estimated partial slopes for the elsewhere aggregates of batters and pitchers
exhibit relative importance on home run intensity that is 
remarkably 
similar to an analysis \citep{abdwr3} based on 2023 data
that uses a logistic regression for home run probabilities 
in individual PAs with random effects for batter and
pitcher.  The square root of the ratio of estimated variance 
components, $\sqrt{\widehat\sigma_B^2/\widehat\sigma_P^2}=2.78$,
is similar to the ratio of personnel effects, 
$\widehat\beta_B/\widehat\beta_P=1.01/0.38=2.66$.

\clearpage 
\begin{table}
\centering
\caption{Ballpark parameter estimates and standard errors. The
reference cell is for games involving right-handed batters and pitchers 
in the ballpark in Washington D.C. ($\widehat\beta^{RR}_{WAS}=0$).} 
\begin{tabular}{c||c|c|c|c} 
&&&& \\ 
Ballpark ($p$)    &     $\widehat\beta_p^{LL}$     &     $\widehat\beta_p^{LR}$ &     $\widehat\beta_p^{RL}$ &     $\widehat\beta_p^{RR}$ \\ \hline
SFN       &      -1.05      &      -0.45      &      -0.35      &      -0.34    \\ 
MIA       &      -1.09      &      -0.33      &      -0.37      &      -0.40    \\ 
PIT       &      -1.19      &      -0.14      &      -0.47      &      -0.39    \\ 
OAK       &      -0.81      &      -0.30      &      -0.32      &      -0.33    \\ 
SLN       &      -1.13      &      -0.19      &      -0.27      &      -0.36    \\ 
KCA       &      -1.05      &      -0.23      &      -0.31      &      -0.27    \\ 
CLE       &      -0.97      &      -0.11      &      -0.40      &      -0.18    \\ 
BOS       &      -0.76      &      -0.27      &      -0.17      &      -0.16    \\ 
SDN       &      -1.03      &      -0.23      &      -0.23      &      -0.11    \\ 
DET       &      -1.03      &      -0.16      &      -0.20      &      -0.19    \\ 
TBA       &      -0.81      &      -0.17      &      -0.26      &      -0.18    \\ 
MIN       &      -1.05      &      -0.20      &      -0.17      &      -0.15    \\ 
ATL       &      -0.91      &      -0.13      &      -0.29      &      -0.15    \\ 
NYN       &      -0.85      &      -0.10      &      -0.23      &      -0.16    \\ 
SEA       &      -0.59      &      -0.16      &      -0.18      &      -0.16    \\ 
ANA       &      -0.66      &      -0.08      &      -0.25      &      -0.15    \\ 
HOU       &      -0.77      &      -0.10      &      -0.18      &      -0.11    \\ 
CHN       &      -0.84      &      -0.16      &      -0.13      &      -0.05    \\ 
ARI       &      -0.86      &      -0.07      &      -0.12      &      -0.08    \\ 
WAS       &      -0.88      &      -0.06      &      -0.25      &       0.00    \\ 
TOR       &      -0.70      &       0.00      &      -0.07      &      -0.07    \\ 
CHA       &      -0.98      &       0.05      &      -0.09      &       0.04    \\ 
LAN       &      -0.50      &       0.01      &      -0.13      &       0.02    \\ 
TEX       &      -0.38      &       0.04      &      -0.15      &       0.01    \\ 
MIL       &      -0.72      &       0.12      &      -0.06      &       0.03    \\ 
BAL       &      -0.67      &       0.16      &      -0.13      &       0.02    \\ 
PHI       &      -0.60      &       0.09      &      -0.08      &       0.08    \\ 
NYA       &      -0.52      &       0.18      &      -0.15      &       0.00    \\ 
COL       &      -0.39      &       0.08      &       0.00      &       0.11    \\ 
CIN       &      -0.62      &       0.24      &      -0.04      &       0.14    \\  \hline
Standard Errors & $(0.07-0.11)$ & $(0.04-0.06)$ & $(0.05-0.06)$ & $(0.04-0.05)$ \\ \hline
\end{tabular}
\label{parms}
\end{table}

Adjusted park means can be constructed by evaluating the inverse link of 
the fitted model, for a given matchup, at average values for the $z$-variables. 
These estimate the mean number of HRs, for a given
matchup, hit at various ballparks if teams with average batting
and pitching HR proficiency at average frequency of plate appearances
were to play at the average frequency for that matchup.
For a ballpark in which observed frequencies differ from league-wide
averages, for example like the aforementioned Progressive Field in Cleveland,
the adjusted means will likely be quite different from
the observed means, as they adjust towards the league average personnel
at average matchup frequency.  
These league-wide averages are summarized
in Table \ref{xvars}.   
The estimated inverse link function evaluated at these averages is
used to obtain the adjusted park means in 
Tables \ref{adjll}, \ref{adjlr}, \ref{adjrl}, and \ref{adjrr}.

\begin{table} [!ht]
\centering
\caption{Average matchup summary.  Average frequencies (counts)
of the four matchups for MLB games, 2010-2023 (excluding 2020).
Also shown are average aggregate elsewhere measures of batter
and pitcher home run proficiency.}
\renewcommand{\arraystretch}{1.4}
\begin{tabular}{cc|c|cc}  
Batter & Pitcher & frequency & $\overline{Z^B}$ & $\overline{Z^P}$ \\ \hline
L & L & 6.6  & 0.14 & 0.15 \\
L & R & 25.8 & 0.75 & 0.76 \\
R & L & 16.3 & 0.50 & 0.50 \\
R & R & 29.1 & 0.84 & 0.86 \\ \hline
\end{tabular} 
\label{xvars} 
\end{table}

Consider the mean number of HRs hit per game
among PAs involving LHBs and RHPs in Yankee Stadium.  
This matchup could take advantage of the short porch 
in right field.  The empirical average among all such
matchups over the 2010-2023 period of observation was $1.025 \ HR/g$. To construct the adjusted
mean at Yankee Stadium,
the
relevant parameter estimates for LR matchups
at Yankee Stadium (ballpark code {\tt NYA}) 
are $\widehat\beta_0 = -1.37$ and 
$\widehat\beta_{NYA}^{LR} = 0.1787$.  
Evaluating the fitted inverse link for 
this ballpark at the average values of 
$z^B=0.75$ and $z^P=0.76$, 
the adjusted mean is
\begin{eqnarray*}
\widehat\lambda_{NYA,LR} &=& 
\exp\{\widehat\beta_0 + \widehat\beta^B (0.75)
+ \widehat\beta^P (0.76) + \widehat\beta_{NYA}^{LR}\} \\ 
&=& \exp\{-1.37 + 1.01(.75) +0.38(0.76) + 0.18\}  \\
&=& 0.875 \ HR/g.  
\end{eqnarray*} 
If Yankee Stadium is conducive to HRs for LR matchups, after controlling
for personnel effects, then this adjusted mean should rank
high compared to other ballparks for LR matchups. Inspection
of Table \ref{adjlr} shows that indeed, it ranks 
second in HR friendliness, being outranked only by the Great
American Ballpark in Cincinnati.  

If Yankee Stadium has, over this period of observation from 2010-2023,
fielded and hosted personnel for which LR matchups are more common 
than in average 
games, and with 
personnel that tend to generate more HRs,
it would be expected that 
adjusting mean $HR/g$ to 
average personnel
would bring it to something below what was observed.  
Indeed that is the 
case here, with the average of the observed aggregate elsewhere HR measure 
for LHB facing RHB in Yankee Stadium taking the value $\overline{z}^{B}_{NYA}=0.7859$.  
For LHP facing RHB, the Home Runs allowed measured elsewhere 
was $\overline{z}^{P}_{NYA}=0.7453$ (see Table \ref{adjlr}).   
These aggregate elsewhere HR measures for batters and pitchers observed
in Yankee Stadium were ranked $7^{th}$ and $18^{th}$ respectively.
The weighted sum of these two explanatory
variables, computed using their respective effect estimates,
$$ Z = \widehat\beta^B Z^B + \widehat\beta^P Z^P $$
constitutes the personnel component of the log HR intensity
of the Poisson model and provides a univariate summary of the 
HR tendency of batters and pitchers.  
For Yankee Stadium, the weighted sum is $z=1.04$ which ranks
$11^{th}$ highest in MLB.
The ranking based on this measure appears along with 
those for the component percentiles in Table \ref{adjlr}.

In the following discussion of the effects of adjustment,
we focus on those ballparks for
which the difference between matchup-specific rank and empirical rank 
are noteworthy.
All changes due to adjustment 
can be analyzed by inspection 
of Tables \ref{adjll}, \ref{adjlr}, \ref{adjrl}, and \ref{adjrr}.

\subsection{Analysis of Adjusted Means}
Matchup-specific inspection of adjusted means reveal which parks are most
and least conducive to home runs.  Some specific observations are given below. \\

\vspace{0.1cm}
\noindent {\em Adjusted means for left-handed batters and left-handed 
pitchers (See Table \ref{adjll})}: \\

For LL matchups, the number of observed $HR/g$ is low, ranging from fewer than
one every 10 games at PNC Park in Pittsburgh ($0.092 \ HR/g$) up to 
$0.233 \ HR/g$ at Globe Field in Texas, 
an amount that is still low, but twice the number at PNC.
If these rates of LL HRs are ranked before and after 
adjustment for personnel effects, 
none of the ballparks have rankings that change very much.
The biggest change is that the ballparks in Kansas City and Detroit swap ranks. 
After adjustment, Kauffman Stadium in Kansas City is the more HR-unfriendly,
ranking $4^{th}$ least friendly, and Comerica Park in Detroit ranks $7^{th}$ 
least friendly.  In Table \ref{adjll}, these ranks appear as $27^{th}$ and $24^{th}$
most HR-friendly. \\

\noindent {\em Adjusted ballpark means for left-handed batters and right-handed pitchers(See Table \ref{adjlr})}: \\

For LR matchups, the magnitude of the adjustment 
is much more pronounced.
The ballparks in Cleveland, Minnesota and the south side of Chicago 
all see double-digit adjustments to their ranks.
Progressive Field in Cleveland hosted 
the highest ranking batters and pitchers, leading to 
$z=1.368$, which is much greater than the
average for LR matchups, $\overline{z}=1.054$.  The 
observed HR rate, $\overline{y}=1.025$,
is greater than any other ballpark in MLB, but the adjusted, $0.657 \ HR/g$ is 
the median, ranked $15$.  Similarly for Target Field in Minnesota, which 
is ranked just behind
Progressive Field for both batters and pitchers, with $z=1.175$.  The observed
$0.771 \ HR/g$, ranked $11$, is adjusted all the way down to $24$.  
PAs in Rate Field, 
home of the Chicago White Sox, involved the 
batters with the lowest ranking and pitchers in the middle,
combining to form the league's lowest HR measure, $z=0.901$.  Upon adjustment for these
low HR PAs, the HR friendliness jumps from unadjusted $0.722 \ HR/g$, ranked $18^{th}$
most friendly, 
up to adjusted $0.767$, ranked $7^{th}$ most HR-friendly.  
These three parks illustrate 
the need to adjust for personnel.  The perception of HR friendliness,
for the LR matchup, changes dramatically as analysis goes from looking only
at matchup-specific empirical HRs to estimates adjusted for batters and pitchers.  \\

\vspace{0.1cm}
\noindent {\em Adjusted ballpark means for right-handed batters and left-handed pitchers(See Table \ref{adjrl})}: \\

For RL matchups, the ballparks in Oakland, Detroit, Philadelphia, Milwaukee and Cincinnati
all see double-digit adjustments in rankings.  Oakland and Detroit are adjusted
way down, as the RL personnel with PAs in those parks are high HR producers, 
especially pitchers, with respective measures of $z=0.819, 0.793$, respectively.
The empirical mean $HR/g$ are ranked 16 and 7, respectively
and adjustment to $\overline{z}$ brings them down to 26 and 17.  Conversely,
Citizen's Bank Ballpark, American Family Field and The Great American Ballpark
in Philadelphia, Milwaukee and Cincinnati, respectively, appear to be among the 
most HR-friendly parks with ranks 5, 3 and 2 after adjustment.  Before adjustment
the rates for RL matchups for these three parks, based only on observed $HR/g$
were 20, 15 and 21. \\

\vspace{0.1cm}
\noindent {\em Adjusted ballpark means for right-handed batters and right-handed pitchers(See Table \ref{adjrr})}: \\

For RR matchups, the Rodgers Centre in Toronto is adjusted from tops in the
league based on unadjusted $1.179 \ HR/g$ down to $0.776$, ranked $12^{th}$. 
The personnel summary was $z=1.431$, with rankings for batters and pitchers 1 and 6, respectively.  
For RR matchups, The Rodgers Centre evidently has 
either seen better HR hitters or RHBs in greater frequency
than any other stadium in baseball, when assessments
are made based on RR matchups by those batters away from the Rodgers Centre.
Conversely, Dodger Stadium and Citizen's Bank Ballpark, in Los Angeles and Philadelphia
respectively, are adjusted up from 21 and 14 to two of the more HR friendly
parks in the league, ranked $6^{th}$ and $3^{rd}$.  RHPs facing RHBs
at Dodger Stadium ranked lowest in baseball at allowing 
HRs (away from Dodger Stadium) with league-low $z^P=0.721$.  Those pitching in Philadelphia were not far behind.

\begin{table} [!ht]
\centering
\caption{Empirical ($HR/g$) and adjusted ($\widehat\lambda_{LL}$) ballpark means and ranks.  
Matchups involve only {\bf left-handed batters and pitchers}.
$z_B$ and $z_P$ denote observed values of personnel covariates
for batters and pitchers, respectively, and $z$ their weighted
sum.  Standard errors of the individual $\widehat\lambda_{LL}$ are
not shown, but the average over 30 ballparks is $0.089$.}
\begin{tabular}{c||c|c|c|c|c|c|c|c|c} 
&&&&&&&&& \\
Park  & $z^B$    & $z^P$    & $z$        & Rank $z$ & $\widehat\lambda_{LL}$ &  rank $\widehat\lambda_{LL}$ 
& $HR/g$  & rank $HR/g$ & $\Delta$ rank \\ \hline
PIT  &  0.117    &  0.119    &  0.164    &   27  &  0.095    &   30   &  0.092  &    30    &     0    \\ 
SLN  &  0.112    &  0.113    &  0.157    &   29  &  0.101    &   29   &  0.097  &    29    &     0    \\ 
MIA  &  0.126    &  0.124    &  0.176    &   24  &  0.105    &   28   &  0.103  &    28    &     0    \\ 
KCA  &  0.142    &  0.165    &  0.207    &   12  &  0.109    &   27   &  0.111  &    24    &    -3    \\ 
MIN  &  0.130    &  0.146    &  0.188    &   17  &  0.109    &   26   &  0.109  &    26    &     0    \\ 
SFN  &  0.160    &  0.177    &  0.230    &    5  &  0.109    &   25   &  0.114  &    23    &    -2    \\ 
DET  &  0.103    &  0.139    &  0.157    &   28  &  0.111    &   24   &  0.107  &    27    &     3    \\ 
SDN  &  0.132    &  0.136    &  0.186    &   19  &  0.112    &   23   &  0.111  &    25    &     2    \\ 
CHA  &  0.128    &  0.130    &  0.180    &   22  &  0.118    &   22   &  0.116  &    21    &    -1    \\ 
CLE  &  0.112    &  0.137    &  0.166    &   26  &  0.118    &   21   &  0.115  &    22    &     1    \\ 
ATL  &  0.155    &  0.143    &  0.212    &    9  &  0.125    &   20   &  0.128  &    20    &     0    \\ 
WAS  &  0.136    &  0.125    &  0.186    &   18  &  0.130    &   19   &  0.129  &    19    &     0    \\ 
ARI  &  0.138    &  0.156    &  0.199    &   14  &  0.132    &   18   &  0.133  &    17    &    -1    \\ 
NYN  &  0.136    &  0.139    &  0.191    &   16  &  0.133    &   17   &  0.133  &    18    &     1    \\ 
CHN  &  0.135    &  0.127    &  0.185    &   20  &  0.135    &   16   &  0.134  &    16    &     0    \\ 
TBA  &  0.152    &  0.151    &  0.213    &    8  &  0.138    &   15   &  0.142  &    14    &    -1    \\ 
OAK  &  0.153    &  0.146    &  0.211    &   10  &  0.139    &   14   &  0.141  &    15    &     1    \\ 
HOU  &  0.143    &  0.127    &  0.194    &   15  &  0.144    &   13   &  0.144  &    13    &     0    \\ 
BOS  &  0.176    &  0.179    &  0.247    &    4  &  0.146    &   12   &  0.155  &    10    &    -2    \\ 
MIL  &  0.100    &  0.117    &  0.146    &   30  &  0.151    &   11   &  0.145  &    12    &     1    \\ 
TOR  &  0.126    &  0.140    &  0.181    &   21  &  0.155    &   10   &  0.153  &    11    &     1    \\ 
BAL  &  0.120    &  0.136    &  0.174    &   25  &  0.160    &    9   &  0.157  &     9    &     0    \\ 
ANA  &  0.148    &  0.137    &  0.203    &   13  &  0.161    &    8   &  0.163  &     8    &     0    \\ 
CIN  &  0.133    &  0.109    &  0.176    &   23  &  0.168    &    7   &  0.165  &     7    &     0    \\ 
PHI  &  0.154    &  0.137    &  0.209    &   11  &  0.171    &    6   &  0.175  &     6    &     0    \\ 
SEA  &  0.210    &  0.213    &  0.294    &    1  &  0.173    &    5   &  0.194  &     4    &    -1    \\ 
NYA  &  0.156    &  0.157    &  0.219    &    7  &  0.186    &    4   &  0.192  &     5    &     1    \\ 
LAN  &  0.193    &  0.177    &  0.264    &    3  &  0.189    &    3   &  0.206  &     3    &     0    \\ 
COL  &  0.157    &  0.179    &  0.228    &    6  &  0.211    &    2   &  0.219  &     2    &     0    \\ 
TEX  &  0.198    &  0.179    &  0.270    &    2  &  0.213    &    1   &  0.233  &     1    &     0    \\  \hline
avg  &  0.143    &  0.145    &  0.200    &       &  0.142    &        &  0.144  &          &          \\ \hline
\end{tabular}
\label{adjll}
\end{table}

\clearpage

\begin{table} [!ht]
\centering
\caption{Empirical ($HR/g$) and adjusted ($\widehat\lambda_{LR}$) ballpark means and ranks.  
Matchups involve only {\bf left-handed batters and right-handed pitchers}.
$z_B$ and $z_P$ denote observed values of personnel covariates
for batters and pitchers, respectively, and $z$ their weighted
sum.  Standard errors of the individual $\widehat\lambda_{LR}$ are
not shown, but the average over 30 ballparks is $0.042$.}
\begin{tabular}{c||c|c|c|c|c|c|c|c|c} 
&&&&&&&&& \\
Park  & $z^B$    & $z^P$    & $z$        & Rank $z$& $\widehat\lambda_{LR}$ &  rank $\widehat\lambda_{LR}$ 
& $HR/g$  & rank $HR/g$ & $\Delta$ rank \\ \hline
SFN  &  0.747    &  0.757    &  1.047    &   17  &  0.468    &   30   &  0.517  &    30    &     0    \\ 
MIA  &  0.698    &  0.735    &  0.989    &   24  &  0.523    &   29   &  0.535  &    29    &     0    \\ 
OAK  &  0.754    &  0.778    &  1.063    &   14  &  0.543    &   28   &  0.619  &    26    &    -2    \\ 
BOS  &  0.840    &  0.782    &  1.151    &    3  &  0.557    &   27   &  0.701  &    21    &    -6    \\ 
SDN  &  0.690    &  0.746    &  0.985    &   25  &  0.581    &   26   &  0.591  &    28    &     2    \\ 
KCA  &  0.746    &  0.812    &  1.067    &   12  &  0.582    &   25   &  0.657  &    24    &    -1    \\ 
MIN  &  0.828    &  0.876    &  1.175    &    2  &  0.600    &   24   &  0.771  &    11    &   -13    \\ 
SLN  &  0.714    &  0.694    &  0.989    &   23  &  0.606    &   23   &  0.605  &    27    &     4    \\ 
TBA  &  0.779    &  0.771    &  1.085    &    9  &  0.617    &   22   &  0.712  &    19    &    -3    \\ 
DET  &  0.693    &  0.753    &  0.991    &   22  &  0.624    &   21   &  0.639  &    25    &     4    \\ 
SEA  &  0.814    &  0.759    &  1.116    &    5  &  0.624    &   20   &  0.759  &    12    &    -8    \\ 
CHN  &  0.761    &  0.739    &  1.055    &   16  &  0.626    &   19   &  0.696  &    23    &     4    \\ 
PIT  &  0.780    &  0.812    &  1.102    &    7  &  0.637    &   18   &  0.746  &    14    &    -4    \\ 
ATL  &  0.800    &  0.778    &  1.109    &    6  &  0.644    &   17   &  0.752  &    13    &    -4    \\ 
CLE  &  0.988    &  0.959    &  1.368    &    1  &  0.657    &   16   &  1.025  &     1    &   -15    \\ 
HOU  &  0.743    &  0.667    &  1.009    &   20  &  0.659    &   15   &  0.699  &    22    &     7    \\ 
NYN  &  0.815    &  0.833    &  1.146    &    4  &  0.665    &   14   &  0.810  &     9    &    -5    \\ 
ANA  &  0.706    &  0.736    &  0.998    &   21  &  0.673    &   13   &  0.706  &    20    &     7    \\ 
ARI  &  0.714    &  0.780    &  1.023    &   19  &  0.679    &   12   &  0.733  &    17    &     5    \\ 
WAS  &  0.772    &  0.793    &  1.086    &    8  &  0.686    &   11   &  0.787  &    10    &    -1    \\ 
TOR  &  0.681    &  0.684    &  0.952    &   28  &  0.733    &   10   &  0.741  &    15    &     5    \\ 
LAN  &  0.678    &  0.644    &  0.934    &   29  &  0.741    &    9   &  0.736  &    16    &     7    \\ 
TEX  &  0.776    &  0.723    &  1.064    &   13  &  0.764    &    8   &  0.877  &     6    &    -2    \\ 
CHA  &  0.625    &  0.700    &  0.901    &   30  &  0.767    &    7   &  0.722  &    18    &    11    \\ 
COL  &  0.687    &  0.724    &  0.974    &   27  &  0.796    &    6   &  0.811  &     8    &     2    \\ 
PHI  &  0.762    &  0.814    &  1.084    &   10  &  0.797    &    5   &  0.902  &     5    &     0    \\ 
MIL  &  0.703    &  0.697    &  0.980    &   26  &  0.822    &    4   &  0.829  &     7    &     3    \\ 
BAL  &  0.761    &  0.745    &  1.056    &   15  &  0.856    &    3   &  0.946  &     4    &     1    \\ 
NYA  &  0.786    &  0.745    &  1.082    &   11  &  0.875    &    2   &  1.025  &     2    &     0    \\ 
CIN  &  0.735    &  0.770    &  1.040    &   18  &  0.926    &    1   &  0.992  &     3    &     2    \\  \hline
Avg   &   0.752  &   0.760  &   1.054  &         &   0.678  &         & 0.755   &            &      \\   \hline
\end{tabular}
\label{adjlr}
\end{table}

\clearpage

\begin{table} [!ht]
\centering
\caption{Empirical ($HR/g$) and adjusted ($\widehat\lambda_{RL}$) ballpark means and ranks.  
Matchups involve only {\bf right-handed batters and left-handed pitchers}.
$z_B$ and $z_P$ denote observed values of personnel covariates
for batters and pitchers, respectively, and $z$ their weighted
sum.  Standard errors of the individual $\widehat\lambda_{RL}$ are
not shown, but the average over 30 ballparks is $0.051$.}
\begin{tabular}{c||c|c|c|c|c|c|c|c|c} 
&&&&&&&&& \\
Park  & $z^B$    & $z^P$    & $z$        & Rank $z$& $\widehat\lambda_{RL}$ &  rank $\widehat\lambda_{RL}$ 
& $HR/g$  & rank $HR/g$ & $\Delta$ rank \\ \hline
PIT  &  0.457    &  0.470    &  0.643    &   22  &  0.319    &   30   &  0.347  &    29    &    -1    \\ 
CLE  &  0.380    &  0.367    &  0.525    &   29  &  0.344    &   29   &  0.326  &    30    &     1    \\ 
MIA  &  0.465    &  0.504    &  0.664    &   20  &  0.354    &   28   &  0.413  &    26    &    -2    \\ 
SFN  &  0.515    &  0.586    &  0.746    &   11  &  0.362    &   27   &  0.447  &    22    &    -5    \\ 
OAK  &  0.581    &  0.601    &  0.819    &    1  &  0.372    &   26   &  0.511  &    16    &   -10    \\ 
KCA  &  0.477    &  0.519    &  0.682    &   18  &  0.374    &   25   &  0.434  &    25    &     0    \\ 
ATL  &  0.419    &  0.409    &  0.581    &   28  &  0.383    &   24   &  0.393  &    28    &     4    \\ 
SLN  &  0.427    &  0.396    &  0.585    &   27  &  0.390    &   23   &  0.409  &    27    &     4    \\ 
TBA  &  0.534    &  0.527    &  0.743    &   12  &  0.396    &   22   &  0.486  &    18    &    -4    \\ 
WAS  &  0.461    &  0.455    &  0.642    &   23  &  0.399    &   21   &  0.443  &    23    &     2    \\ 
ANA  &  0.572    &  0.561    &  0.794    &    3  &  0.399    &   20   &  0.529  &    12    &    -8    \\ 
SDN  &  0.499    &  0.554    &  0.719    &   14  &  0.406    &   19   &  0.500  &    17    &    -2    \\ 
NYN  &  0.434    &  0.441    &  0.609    &   25  &  0.406    &   18   &  0.435  &    24    &     6    \\ 
DET  &  0.560    &  0.588    &  0.793    &    5  &  0.419    &   17   &  0.563  &     7    &   -10    \\ 
HOU  &  0.519    &  0.517    &  0.724    &   13  &  0.426    &   16   &  0.515  &    14    &    -2    \\ 
SEA  &  0.557    &  0.598    &  0.794    &    4  &  0.427    &   15   &  0.562  &     8    &    -7    \\ 
MIN  &  0.457    &  0.476    &  0.646    &   21  &  0.429    &   14   &  0.485  &    19    &     5    \\ 
BOS  &  0.557    &  0.525    &  0.765    &    9  &  0.430    &   13   &  0.551  &     9    &    -4    \\ 
TEX  &  0.562    &  0.592    &  0.796    &    2  &  0.439    &   12   &  0.577  &     5    &    -7    \\ 
NYA  &  0.582    &  0.523    &  0.791    &    6  &  0.442    &   11   &  0.600  &     2    &    -9    \\ 
LAN  &  0.568    &  0.519    &  0.775    &    8  &  0.447    &   10   &  0.564  &     6    &    -4    \\ 
CHN  &  0.488    &  0.481    &  0.679    &   19  &  0.451    &    9   &  0.517  &    13    &     4    \\ 
BAL  &  0.491    &  0.500    &  0.689    &   17  &  0.451    &    8   &  0.533  &    11    &     3    \\ 
ARI  &  0.505    &  0.528    &  0.715    &   15  &  0.454    &    7   &  0.548  &    10    &     3    \\ 
CHA  &  0.546    &  0.540    &  0.760    &   10  &  0.468    &    6   &  0.597  &     4    &    -2    \\ 
PHI  &  0.416    &  0.441    &  0.591    &   26  &  0.472    &    5   &  0.479  &    20    &    15    \\ 
TOR  &  0.571    &  0.520    &  0.778    &    7  &  0.477    &    4   &  0.634  &     1    &    -3    \\ 
MIL  &  0.449    &  0.423    &  0.617    &   24  &  0.482    &    3   &  0.513  &    15    &    12    \\ 
CIN  &  0.345    &  0.395    &  0.501    &   30  &  0.493    &    2   &  0.449  &    21    &    19    \\ 
COL  &  0.494    &  0.530    &  0.704    &   16  &  0.511    &    1   &  0.599  &     3    &     2    \\  \hline
Avg   &   0.496  &   0.503  &   0.696  &         &   0.421  &         & 0.499   &            &         \\ \hline
\end{tabular}
\label{adjrl}
\end{table}

\clearpage

\begin{table} [!ht]
\centering
\caption{Empirical ($HR/g$) and adjusted ($\widehat\lambda_{RR}$) 
ballpark means and ranks.  
Matchups involve only {\bf right-handed batters and pitchers.}
$z_B$ and $z_P$ denote observed values of personnel covariates
for batters and pitchers, respectively, and $z$ their weighted
sum.  Standard errors of the individual $\widehat\lambda_{RR}$ are
not shown, but the average over 30 ballparks is $0.041$.}
\begin{tabular}{c||c|c|c|c|c|c|c|c|c} 
&&&&&&&&& \\
Park  & $z^B$    & $z^P$    & $z$        & Rank $z$& $\widehat\lambda_{RR}$ &  rank $\widehat\lambda_{RR}$ 
& $HR/g$  & rank $HR/g$ & $\Delta$ rank \\ \hline
MIA  &  0.889    &  0.965    &  1.270    &    7  &  0.555    &   30   &  0.680  &    26    &    -4    \\ 
PIT  &  0.775    &  0.906    &  1.133    &   22  &  0.565    &   29   &  0.590  &    29    &     0    \\ 
SLN  &  0.986    &  0.972    &  1.372    &    2  &  0.582    &   28   &  0.794  &    19    &    -9    \\ 
SFN  &  0.683    &  0.747    &  0.979    &   30  &  0.589    &   27   &  0.523  &    30    &     3    \\ 
OAK  &  0.861    &  0.829    &  1.190    &   13  &  0.600    &   26   &  0.731  &    23    &    -3    \\ 
KCA  &  0.798    &  0.873    &  1.144    &   21  &  0.634    &   25   &  0.674  &    27    &     2    \\ 
DET  &  0.879    &  0.893    &  1.233    &   10  &  0.688    &   24   &  0.811  &    18    &    -6    \\ 
TBA  &  0.821    &  0.846    &  1.157    &   18  &  0.695    &   23   &  0.761  &    22    &    -1    \\ 
CLE  &  0.750    &  0.782    &  1.060    &   26  &  0.697    &   22   &  0.669  &    28    &     6    \\ 
SEA  &  0.762    &  0.724    &  1.049    &   27  &  0.707    &   21   &  0.710  &    25    &     4    \\ 
BOS  &  0.870    &  0.850    &  1.208    &   11  &  0.708    &   20   &  0.833  &    15    &    -5    \\ 
NYN  &  0.783    &  0.789    &  1.096    &   24  &  0.711    &   19   &  0.721  &    24    &     5    \\ 
MIN  &  0.902    &  0.875    &  1.250    &    9  &  0.713    &   18   &  0.875  &    12    &    -6    \\ 
ANA  &  0.953    &  0.954    &  1.331    &    4  &  0.716    &   17   &  0.969  &     9    &    -8    \\ 
ATL  &  0.845    &  0.875    &  1.192    &   12  &  0.717    &   16   &  0.813  &    17    &     1    \\ 
SDN  &  0.792    &  0.831    &  1.121    &   23  &  0.743    &   15   &  0.786  &    20    &     5    \\ 
HOU  &  0.960    &  0.960    &  1.341    &    3  &  0.744    &   14   &  1.020  &     5    &    -9    \\ 
ARI  &  0.799    &  0.882    &  1.148    &   20  &  0.770    &   13   &  0.830  &    16    &     3    \\ 
TOR  &  1.058    &  0.936    &  1.431    &    1  &  0.776    &   12   &  1.179  &     1    &   -11    \\ 
CHN  &  0.836    &  0.860    &  1.178    &   16  &  0.787    &   11   &  0.876  &    11    &     0    \\ 
WAS  &  0.812    &  0.879    &  1.160    &   17  &  0.831    &   10   &  0.903  &    10    &     0    \\ 
NYA  &  0.862    &  0.799    &  1.180    &   15  &  0.835    &    9   &  0.990  &     8    &    -1    \\ 
TEX  &  0.776    &  0.800    &  1.093    &   25  &  0.838    &    8   &  0.855  &    13    &     5    \\ 
BAL  &  0.942    &  0.927    &  1.310    &    5  &  0.848    &    7   &  1.126  &     2    &    -5    \\ 
LAN  &  0.697    &  0.721    &  0.982    &   29  &  0.848    &    6   &  0.763  &    21    &    15    \\ 
MIL  &  0.924    &  0.928    &  1.292    &    6  &  0.857    &    5   &  1.051  &     4    &    -1    \\ 
CHA  &  0.857    &  0.830    &  1.187    &   14  &  0.866    &    4   &  0.998  &     6    &     2    \\ 
PHI  &  0.703    &  0.769    &  1.007    &   28  &  0.896    &    3   &  0.841  &    14    &    11    \\ 
COL  &  0.808    &  0.861    &  1.150    &   19  &  0.928    &    2   &  0.995  &     7    &     5    \\ 
CIN  &  0.876    &  0.952    &  1.253    &    8  &  0.961    &    1   &  1.114  &     3    &     2    \\  \hline
Avg   &   0.842  &   0.861  &   1.183  &         &   0.747  &   & 0.849   &            &         \\ \hline
\end{tabular}
\label{adjrr}
\end{table}

\subsection{Marginal adjusted ballpark means}
While it is clear that ballpark HR frequencies depend strongly on handedness,
there may still be interest in an overall assessment of HR friendliness
that averages over matchups.  One such measure is the marginal adjusted mean,
or the simple sum of adjusted means, on the inverse link scale.  
If the adjusted mean for park $p$ and matchup $BP$ is denoted $\widehat\lambda_{p,BP}$,
then the marginal adjusted mean is defined as
$$ \widehat\lambda_p =  \widehat\lambda_{p,LL} + \widehat\lambda_{p,LR} + \widehat\lambda_{p,RL} + \widehat\lambda_{p,RR}. $$
This construct appears along with the matchup-specific adjusted
means in Table \ref{overall}.
As an example, consider Yankee Stadium, 
with adjusted means for the four matchups given below: 

\begin{eqnarray*}
\widehat\lambda_{NYA,\ LL} &=& \exp\{\widehat\beta_0 + \widehat\beta^B(0.14) + \widehat\beta^P(0.15) + \widehat\beta^{LL}_{NYA}\} = 0.186 \\
\widehat\lambda_{NYA,\ LR} &=& \exp\{\widehat\beta_0 + \widehat\beta^B(0.75) + \widehat\beta^P(0.76) + \widehat\beta^{LR}_{NYA}\} =0.875 \\
\widehat\lambda_{NYA,\ RL} &=& \exp\{\widehat\beta_0 + \widehat\beta^B(0.50) + \widehat\beta^P(0.50) + \widehat\beta^{RL}_{NYA}\} =0.442 \\
\widehat\lambda_{NYA,\ RR} &=& \exp\{\widehat\beta_0 + \widehat\beta^B(0.84) + \widehat\beta^P(0.86) + \widehat\beta^{RR}_{NYA}\} =0.835.
\end{eqnarray*}
The marginal adjusted mean for Yankee Stadium is then just the sum
$$\widehat\lambda_{NYA} = 0.186 + 0.875 + 0.442 + 0.835 = 2.34.$$
No special weights are required because the frequency associated
with the matchups is baked into the aggregate batter and pitcher measures
of home run proficiency, $\overline{Z^B}$ and $\overline{Z^P}$ for
the four matchups.
If Yankee Stadium were to host two teams that are average with
respect to matchup frequency and player HR ability, the estimated
marginal mean number of home runs in a population of such games is 2.34 $HR/g$.
The observed mean was 2.74, with the difference somewhat attributable
to the elevated levels of $z^B$ and $z^P$ at the ballpark. 
$z^B$ is above the median for all matchups and $z^P$ is below the 
median only 
for RHP.

The empirical cumulative distribution function (ECDF) of the marginal adjusted means in the column titled $\widehat\lambda$ from 
Table \ref{overall} 
is plotted
in Figure 3
with selected teams emphasized with text in the plot.  Ballparks have been informally clustered at values for the 
overall adjusted mean where the ECDF appears to take larger jumps, and these five resulting cluster labels also
appear in Table \ref{overall}.  Immediate observations from inspection of the table and plot may be unsurprising to baseball
fans. Ballparks in Cluster 1 have reputations for being pitcher-friendly. 
Cluster 2 contains the ballparks in Cleveland and Boston, both of which rank in the middle in empirical
HRs seen ($19^{th}$ and $15^{th}$, respectively), but are estimated to be considerably less HR-friendly  after
adjustment (ranked $24^{th}$ and $23^{rd}$, respectively).  The ballpark in Toronto lands in Cluster 3 and observed 
the 4th most HRs among all ballparks, but ranks only 10th most friendly after adjustment. 
This cluster also contains
Yankee Stadium, where more HRs were hit per game than in any other ballpark, but which ranks third after adjustment, 
behind the two ballparks that analysis indicates are the most HR-friendly in baseball in Cincinnati and Colorado.  It is 
interesting also to note that the Great American Ballpark appears to be in a league of its own according to the 
marginal adjusted mean.

\begin{table} [!ht]
\centering
\caption{Marginal adjusted ballpark means.  
Table ordered according to marginal adjusted mean home runs, denoted
by $\widehat\lambda$. $HR/g$ gives empirical ballpark home runs per game (with
a column for Rank).  
The $\Delta$ Rank column indicates the change in rank due to adjustment.}
\begin{tabular}{c|c||c|c|c|c|c|c|c|c|c}  
&&&&&&&&&& \\
Cluster & Park & LL  &  LR    &  RL    &  RR    &$\widehat\lambda$ & Rank $\widehat\lambda$ & $HR/g$ & Rank $HR/g$ &  $\Delta$ Rank      \\ \hline
 1  &  SFN  &  0.11  &  0.47  &  0.36  &  0.59  &  1.53  &  30  &  1.57  &  30  &   0  \\
    &  MIA  &  0.11  &  0.52  &  0.35  &  0.56  &  1.54  &  29  &  1.68  &  29  &   0  \\
    &  PIT  &  0.09  &  0.64  &  0.32  &  0.57  &  1.62  &  28  &  1.74  &  28  &   0  \\
    &  OAK  &  0.14  &  0.54  &  0.37  &  0.60  &  1.65  &  27  &  1.96  &  24  &  -3  \\
    &  SLN  &  0.10  &  0.61  &  0.39  &  0.58  &  1.68  &  26  &  1.85  &  26  &   0  \\
    &  KCA  &  0.11  &  0.58  &  0.37  &  0.63  &  1.70  &  25  &  1.83  &  27  &   2  \\ \hdashline
 2  &  CLE  &  0.12  &  0.66  &  0.34  &  0.70  &  1.82  &  24  &  2.09  &  19  &  -5  \\
    &  BOS  &  0.15  &  0.56  &  0.43  &  0.71  &  1.84  &  23  &  2.20  &  15  &  -8  \\
    &  SDN  &  0.11  &  0.58  &  0.41  &  0.74  &  1.84  &  22  &  1.94  &  25  &   3  \\
    &  DET  &  0.11  &  0.62  &  0.42  &  0.69  &  1.84  &  21  &  2.06  &  20  &  -1  \\
    &  TBA  &  0.14  &  0.62  &  0.40  &  0.69  &  1.85  &  20  &  2.05  &  22  &   2  \\
    &  MIN  &  0.11  &  0.60  &  0.43  &  0.71  &  1.85  &  19  &  2.20  &  14  &  -5  \\
    &  ATL  &  0.13  &  0.64  &  0.38  &  0.72  &  1.87  &  18  &  2.05  &  21  &   3  \\
    &  NYN  &  0.13  &  0.66  &  0.41  &  0.71  &  1.91  &  17  &  2.04  &  23  &   6  \\
    &  SEA  &  0.17  &  0.62  &  0.43  &  0.71  &  1.93  &  16  &  2.17  &  18  &   2  \\
    &  ANA  &  0.16  &  0.67  &  0.40  &  0.72  &  1.95  &  15  &  2.28  &  11  &  -4  \\
    &  HOU  &  0.14  &  0.66  &  0.43  &  0.74  &  1.97  &  14  &  2.29  &  10  &  -4  \\
    &  CHN  &  0.13  &  0.63  &  0.45  &  0.79  &  2.00  &  13  &  2.18  &  17  &   4  \\
    &  ARI  &  0.13  &  0.68  &  0.45  &  0.77  &  2.04  &  12  &  2.18  &  16  &   4  \\
    &  WAS  &  0.13  &  0.69  &  0.40  &  0.83  &  2.05  &  11  &  2.21  &  13  &   2  \\ \hdashline
 3  &  TOR  &  0.15  &  0.73  &  0.48  &  0.78  &  2.14  &  10  &  2.64  &   4  &  -6  \\
    &  CHA  &  0.12  &  0.77  &  0.47  &  0.87  &  2.22  &   9  &  2.38  &   8  &  -1  \\
    &  LAN  &  0.19  &  0.74  &  0.45  &  0.85  &  2.23  &   8  &  2.23  &  12  &   4  \\
    &  TEX  &  0.21  &  0.76  &  0.44  &  0.84  &  2.25  &   7  &  2.50  &   6  &  -1  \\
    &  MIL  &  0.15  &  0.82  &  0.48  &  0.86  &  2.31  &   6  &  2.46  &   7  &   1  \\
    &  BAL  &  0.16  &  0.86  &  0.45  &  0.85  &  2.32  &   5  &  2.71  &   2  &  -3  \\
    &  PHI  &  0.17  &  0.80  &  0.47  &  0.90  &  2.34  &   4  &  2.35  &   9  &   5  \\
    &  NYA  &  0.19  &  0.87  &  0.44  &  0.84  &  2.34  &   3  &  2.74  &   1  &  -2  \\ \hdashline
 4  &  COL  &  0.21  &  0.80  &  0.51  &  0.93  &  2.45  &   2  &  2.56  &   5  &   3  \\ \hdashline
 5  &  CIN  &  0.17  &  0.93  &  0.49  &  0.96  &  2.55  &   1  &  2.66  &   3  &   2  \\ \hline
\end{tabular}
\label{overall}
\end{table}

\begin{center}
Figure 3 about here
\end{center}

\subsection{Variance explained}
To quantify the amount of variability 
in the observed number of HRs hit per game, the model 
is
compared with 
several generalized linear mixed models containing subsets of the
explanatory variables included in the full model.
In these models, the response analyzed is the difference between observed home
runs and the amount predicted by the generalized linear mixed model, 
\begin{eqnarray*}
\widehat\lambda_g &=& \widehat\lambda^{LL}(z^B_g,z^P_g) + 
\widehat\lambda^{LR}(z^B_g,z^P_g) + 
\widehat\lambda^{RL}(z^B_g,z^P_g) +  
\widehat\lambda^{RR}(z^B_g,z^P_g)  \\
r_g&=&y_g - \widehat\lambda_g
\end{eqnarray*}
for game $g=1,\ldots,32325.$  Here it is emphasized that the observed
explanatory variables $z^B_g,z^P_g$ are aggregated only over the
plate appearances for the corresponding matchup for game $g$.  
If a matchup 
is
not observed for a game, as would be the case
in a duel of two RHPs who each threw complete games, the contribution
of 0 to $\widehat\lambda_g$ is adopted.

Candidate models
to obtain the predictions $\widehat\lambda_g$ include 
subsets of
the following terms: random season effects, factorial park-by-matchup 
effects, the elsewhere batting covariate $Z^B$, and the elsewhere 
pitching covariate, $Z^P$.  For each model, the variance of the 
residuals, $s^2$ was computed and these variances are given in
Table \ref{varexplain}.

\begin{table} [!ht]
\centering
\caption{Residual variance, $s^2$ and AIC for subset models 
for game home run totals.
$Z^B$ and $Z^P$ are elsewhere measures of batter and pitcher
tendendies 
to hit and allow home runs, respectively.  }
\begin{tabular}{llll}
Model & $df$ & $s^2$ & AIC \\ \hline
Full & 32 & 2.42 & 220189.0\\
season, Park-by-matchup & 30& 2.50 & 237693.4\\
season, $Z^B, Z^P$ & 3 & 2.58 & 223271.2 \\
season, $Z^B$      & 2 & 2.60 & 224061.8 \\
season, $Z^p$      & 2 & 4.07 & 229402.0 \\ \hline
\end{tabular}
\label{varexplain}
\end{table}

Inspection of the reduction in $s^2$ with models of increasing complexity
suggests that accounting for batter ability and frequency
appears to be more important than accounting for pitcher 
ability and frequency.  
This result is consistent with the relative magnitude of the slope
estimates for batter and pitcher in the full model and with the
estimated variance components in the logistic regression model
for 2023 data \citep{abdwr3} mentioned in Section 3.
The model with park-matchup effects, 
but without the aggregate measures of batter and pitcher ability, yields
a smaller $s^2$ than 
the models with only ability covariates.
Adding those covariates to the model with park-by-matchup effects
brings about a modest 
improvement in variance explained.

\subsection{Division effects}

Except for 2023, the data analyzed here were collected before MLB introduced the balanced
schedule, which has been shown to affect competitive balance in baseball \cite{lee}.  In 2010-2022, teams played more 
games against teams within their division.
This may result in greater imbalance in personnel across ballparks if there is variation
in batting or pitching ability by division.  To investigate this possibility, linear models with factorial
effects of division were fit, using the magnitude of the adjustment for each park, division
and matchup as a response:
$$ A_{ij} = \widehat\lambda(\overline{Z}^B,\overline{Z}^P) - \overline{y}_{ij},$$
where $i=1,\ldots,6$ is an index for division and $j$ is an index for 
for ballpark.  
Separate factorial effects models of
the form $A_{ij} = \mu + \alpha_i + \epsilon_{ij}$  were fit
for each of the four matchups. The relative magnitudes of the division sum of 
squares from the analysis of variance indicate much of the adjustment that this model 
quantifies can be attributed to what division a ballpark is in.  
Table \ref{divisions} lists the teams within each division 
along with their three letter abbreviations and Table \ref{anova}
shows the decomposition of total sums of squares in the factorial
effects models.

\begin{table} [!ht]
\centering
\caption{Listing of all 30 teams and their divisions, along with three
letter abbreviations.}
\begin{tabular}{ c c c } 
\multicolumn{3}{ c }{National League} \\
West & Central & East \\ \hline 
ARI:Arizona Diamondbacks & CHN:Chicago Cubs        & ATL:Atlanta Braves \\ 
COL:Colorado Rockies     & CIN:Cincinnati Reds     & MIA:Miami Marlins \\
LAN:Los Angeles Dodgers  & MIL:Milwaukee Brewers   & NYN:New York Mets \\
SDN:San Diego Padres     & PIT:Pittsburgh Pirates  & PHI:Philadelphia Phillies\\
SFN:San Francisco Giants & SLN:ST. Louis Cardinals & WAS:Washington Nationals \\ \hline
\multicolumn{3}{ c }{American League} \\
West & Central & East \\ \hline 
HOU:Houston Astros     & DET:Detroit Tigers      & BAL:Baltimore Orioles \\
LAA:Los Angeles Angels & CHA:Chicago White Sox   & BOS:Boston Red Sox \\
OAK:Oakland Athletics  & CLE:Cleveland Guardians & NYA:New York Yankees \\
SEA:Seattle Mariners   & KCA:Kansas City Royals  & TBA:Tampa Bay Rays \\
TEX:Texas Rangers      & MIN:Minnesota Twins     & TOR:Toronto Blue Jays \\ \hline 
\end{tabular}
\label{divisions}
\end{table}

\begin{table} [!ht]
\centering
\caption{Analysis of variance of division effects on the magnitude
of the adjustment. }
\begin{tabular}{ c|c|c|c } 
Matchup & $SS[\mbox{Division}]$ & $SS[\mbox{Tot}]$ & $r^2$ \\
& $df=5$ & $df=24$ & $=SS[\mbox{division}]/SS[\mbox{Tot}]$ \\ \hline
  LL    &  0.0006 &  0.0014   &  0.42 \\
  LR    &  0.0290 &  0.1753   &  0.17 \\
  RL    &  0.0514 &  0.0843   &  0.61 \\
  RR    &  0.1275 &  0.3559   &  0.36 \\ \hline
\end{tabular}
\label{anova}
\end{table}

\section{Conclusion}
Because of the considerable differences among ballparks in MLB, their 
effect on play has long held the public's interest.  Many approaches have 
been developed to quantify these effects, but not without considerable 
criticism.  In particular, due to asymmetries, these effects should take 
batter and/or pitcher handedness into account.  A model was developed 
that takes the number of HRs hit in a game to behave like a Poisson 
random variable, conditionally on matchup handedness, ballpark, season 
and personnel covariates.  Poissonness plots \citep{hoaglin}
suggest that the 
model provides a reasonable fit to data collected over the 2010-2023 
period, excluding the Covid season.  Inspection of HR frequencies 
indicates ballpark effects of considerable magnitude.  In the case 
of LL matchups, both empirical and adjusted mean HR frequencies for 
Globe Field in Texas are more than twice as large as those for PNC Park 
in Pittsburgh.  For a given ballpark, the relative ranking can be quite 
different for one matchup than it is for another.  
In some
cases,  the ranks of means adjusted for personnel are dramatically
different than ranks based only on empirical means that do not
control for personnel differences.  RL matchups at the Great American 
Ballpark in Cincinnati or Citizen's Ballpark in Philadelphia are 
two examples.  The methodology reveals parks that exhibit 
asymmetric HR frequencies.  Yankee Stadium in New York is an example, with
means for LHB 
against
RHP and LHP ranked $2^{nd}$ and $4^{th}$, respectively, while the 
corresponding ranks
for RHBs are $11^{th}$ and $13^{th}$; a finding that is consistent with 
the ballpark's
reputation based on the ``short porch'' in right field.
Similar advantages in adjusted means for RHBs are observed for Wrigley
Field in Chicago and Petco Park in San Diego.

The construction of these adjusted means could be considered an 
analysis of covariance, with the personnel measures for batter and 
pitcher tendencies to hit and
allow HRs as 
covariates.  For a particular game-matchup 
observation, $g$, these elsewhere covariates were constructed from HR 
frequencies observed at all other parks except
the one involved in observation $g$.  Addition of these covariates 
to a prediction model that includes random season effects, and 
fixed handedness-by-ballpark
interaction effects brings about
reduced AIC and residual variance.  
Further, by estimating the inverse link function at average values 
of the personnel covariates, ballparks can
be compared on an even footing, with the considerable differences 
among teams, and games, and players within games taken into account.  
Estimating the standard errors
of these adjusted means is straightforward, and offers an important additional inference beyond simple univariate summaries like the 
HRPF.
With the Poisson assumption, these adjusted means also serve as estimates for the population variance, facilitating quantification of the entire distribution of home runs for a game-matchup combination.

The personnel adjustment for a ballpark is 
defined here as the difference 
between 
the empirical mean HR frequency over all games and the mean adjusted to 
leaguewide averages of the personnel covariates.  
Using general linear models for these matchup-specific personnel adjustments
to home run frequencies, analysis of variance is 
used to quantify
the effect of the MLB division in which a team plays.
The coefficients of determination for the four matchups are 
$r^2_{LR}=0.17,r^2_{RR}=0.36, r^2_{LL}=0.42, r^2_{RL}=0.61$.  This 
evidence of 
division effects highlights the importance of adjustments for 
the personnel who tend to have PAs at each park.

A similar approach could be used for other outcomes that are based on
counts, such as earned runs or extra-base hits, or even flyouts or 
foul-outs.  Preliminary analyses not shown here were able to identify
certain parks as producing higher adjusted mean foul-outs than others,
and could also assist in the selection of certain types of personnel,
groundball pitchers over flyball pitchers, for example. 

An interesting next step would be to adjust estimates of the abilities of
individual batters to hit and pitchers to prevent HRs that take handedness
and quality of the opponents, ballpark and division into account.  
This could improve the 
assessment of how players might perform with a change of scenery.  It could 
be used to assist decisions about whom to aquire and whom to release. 
Another
interesting aspect to ballpark effects is whether and how they change over
time, beyond other apparent league-wide trends in scoring and home runs.

\newpage

{\large {\bf Figures:}}
\begin{figure}
\includegraphics[scale=.8]{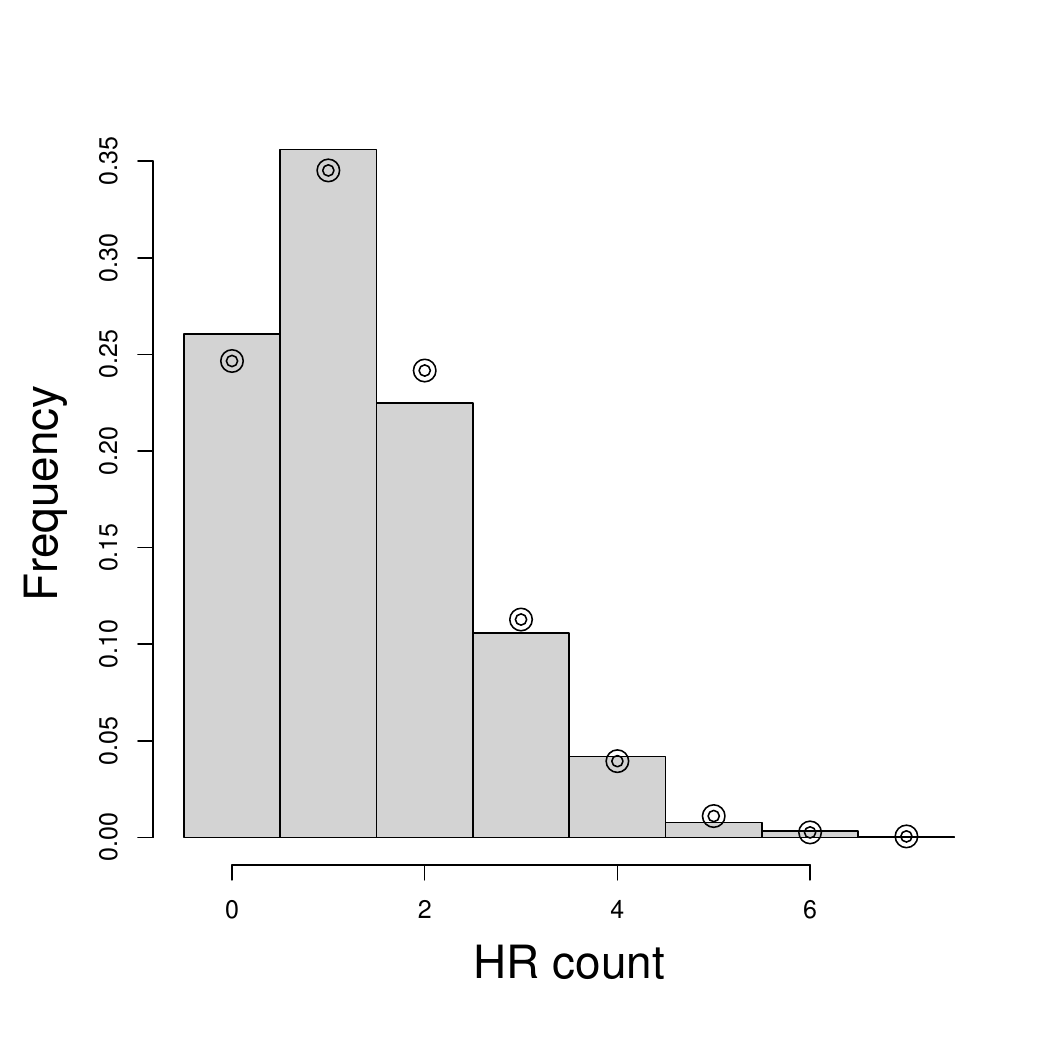}
\caption{Probability histogram for empirical frequencies of
home run totals for 3132 game-matchups with rounded 
estimated expectancy $\widehat\lambda=1.4$.  Also plotted
with dots is the estimated Poisson probability mass function 
with $\widehat\lambda=1.4$.}
\end{figure}

\begin{figure}
\includegraphics[scale=.8]{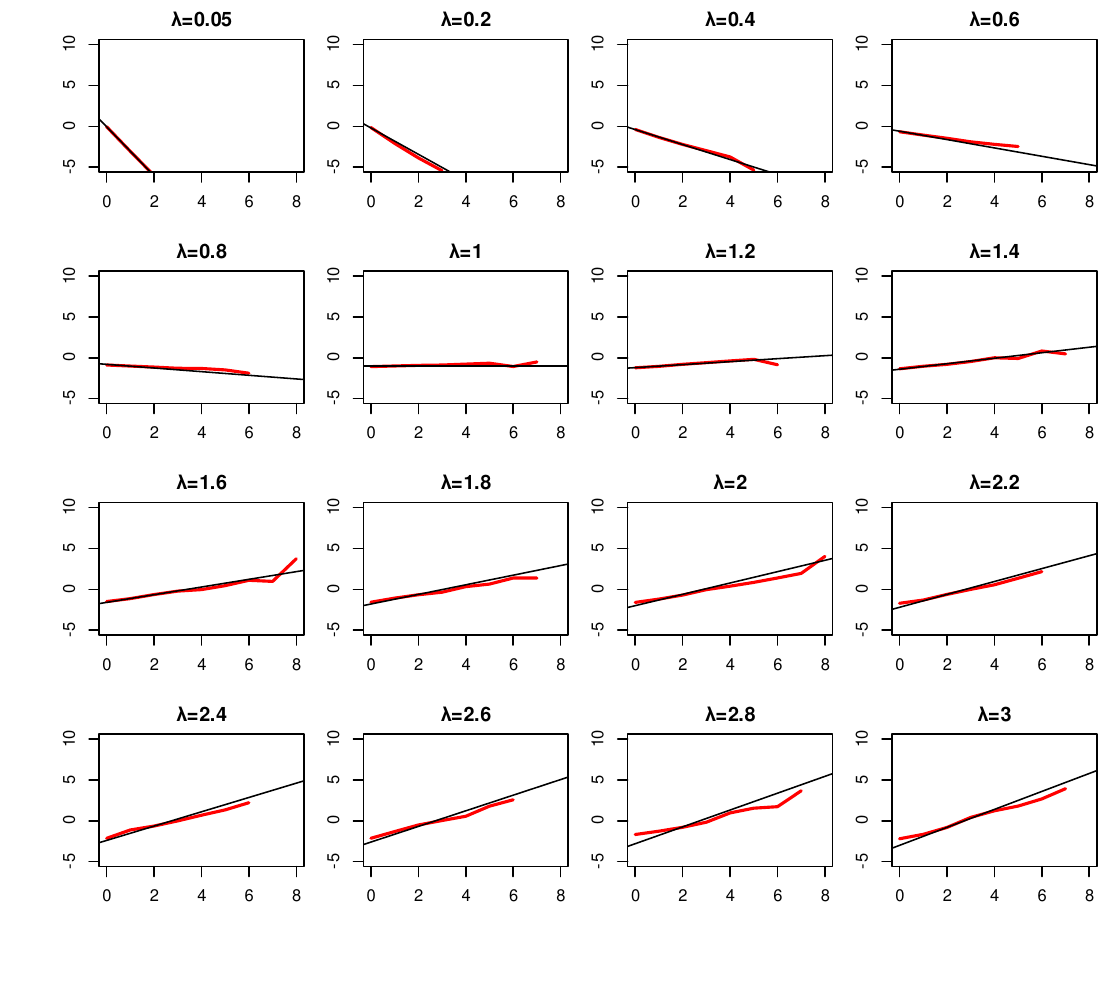}
\caption{Poissonness plots of $\phi(k)=\log(k!n_k/N)$ against
$k$, where $n_k$ is the number of game-matchups with $k$
observed home runs, $N$ is the total number of game-matchups
for which $\widehat\lambda$, rounded to the nearest $0.2$,
takes the given value.  Red is the observed plot, black
is the theoretical line, with intercept $-\widehat\lambda$ and slope 
equal to $-\log(\widehat\lambda)$}
\end{figure}

\begin{figure}
\includegraphics[scale=.8]{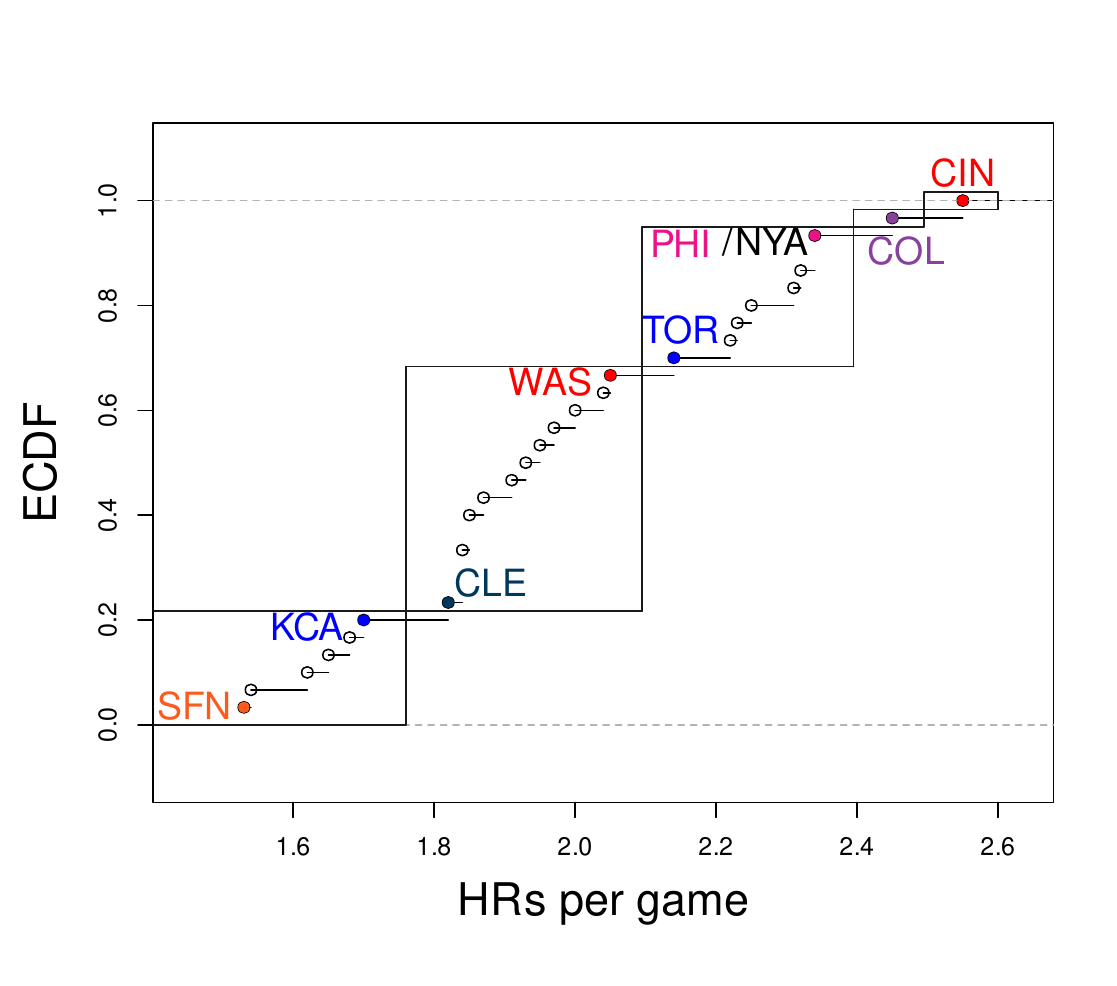}
\caption{Empirical cumulative distribution function of adjusted
overall park means.  Ballparks have been clustered,
depicted in boxes, with those at boundaries of the 
clusters annotated with three-character team/ballpark
abbreviations and colors. }
\end{figure}
\end{document}